\documentstyle[12pt]{article}
\newlength{\extraspace}
\newlength{\extraspaces}
\newcommand{\be}{\begin{equation}
\addtolength{\abovedisplayskip}{\extraspaces}
\addtolength{\belowdisplayskip}{\extraspaces}
\addtolength{\abovedisplayshortskip}{\extraspace}
\addtolength{\belowdisplayshortskip}{\extraspace}}
\newcommand{\ee}{\end{equation}}
\newcommand{\ba}{\begin{eqnarray}
\addtolength{\abovedisplayskip}{\extraspaces}
\addtolength{\belowdisplayskip}{\extraspaces}
\addtolength{\abovedisplayshortskip}{\extraspace}
\addtolength{\belowdisplayshortskip}{\extraspace}}
\newcommand{\ea}{\end{eqnarray}}
\newcommand{\nonu}{\nonumber \\[.5mm]}
\newcommand{\A}{&\!\!\!}

\newcommand{\R}{\bf R}

\newcommand{\e}{\, {\rm e}}
\newcommand{\D}{{\cal D}}

\begin{document}
\thispagestyle{empty}
\begin{flushright}
STUPP--92--130 \\
August, 1992
\end{flushright}
\vspace{15mm}
\begin{center}
{\large{\bf{Semiclassical Quantization of \\[2mm]
Two-Dimensional Dilaton Gravity}}} \\[25mm]
{\sc Yoshiaki Tanii} \\[15mm]
{\it Physics Department, Saitama University \\[2mm]
Urawa, Saitama 338, Japan} \\[25mm]
{\bf Abstract}\\[1cm]
{\parbox{13cm}{\hspace{5mm}
Quantization of the dilaton gravity in two dimensions is discussed
by a semiclassical approximation. We compute the fixed-area
partition function to one-loop order and obtain the string
susceptibility on Riemann surfaces of arbitrary genus. Our result
is consistent with the approach using techniques of conformal
field theories.}}
\end{center}
\vfill
\newpage
\setcounter{section}{0}
\setcounter{equation}{0}
\addtolength{\baselineskip}{2mm}
%
%
A two-dimensional metric-dilaton system coupled to matter fields
(dilaton gravity) was proposed in refs.\ \cite{CGHS,VERLINDE}
as a simple model to discuss the quantum theory of black holes.
Although this model is exactly solvable at the classical level,
the quantum theory is not yet fully understood.
\par
Recently, quantization of this model was discussed in
refs.\ \cite{STR,ABC} using techniques of conformal field theories.
(Quantization was also discussed using other methods in ref.\
\cite{RT}.) The authors in refs.\ \cite{STR,ABC} made an ansatz
about the functional measures of path integrals following a
procedure applied to ordinary two-dimensional gravity in
ref.\ \cite{DDK}.
In the case of ordinary gravity the ansatz was justified by
comparing its results with those of the matrix models \cite{KM}.
The ansatz was also checked by other approaches such as
semiclassical analyses \cite{ZAM,CKT}, the light-cone gauge
quantization \cite{KPZ} and direct calculations of the functional
measures \cite{MMDK}. Since matrix models for the dilaton gravity
are not known at present, it is important to study other approaches
and compare their results with those of refs.\ \cite{STR,ABC}.
\par
The purpose of the present paper is to study the dilaton gravity
by a semiclassical approximation, which becomes exact for a matter
central charge $c \rightarrow - \infty$.
We find a saddle point of the path integral with a fixed `area' and
quantize fluctuations of fields around it to one-loop order.
In particular, we compute the string susceptibility on Riemann
surfaces of arbitrary genus. Our result is consistent with that of
refs.\ \cite{STR,ABC}. We also study a case in which the functional
measures are modified as in ref.\ \cite{STR}.
\par
%
%
We consider a conformal field theory with a central charge $c$
coupled to a metric-dilaton system on a compact closed surface.
The metric is chosen to have the Euclidean signature.
The classical action is
\be
S = {1 \over 2\pi} \int d^2\xi \sqrt{g} \e^{-2\phi}
\left( R_g + 2\mu + 4 g^{\alpha\beta} \partial_\alpha \phi
\partial_\beta \phi \right) + S_{\rm M}[g_{\alpha\beta}, f],
\label{action}
\ee
where $\phi$ is the dilaton field, $R_g$ is the scalar curvature of
the metric $g_{\alpha\beta}$ and $S_{\rm M}$ is the action for the
matter field $f$. The parameter $\mu$ is a generalized cosmological
constant. We have chosen the sign of the
metric-dilaton terms in eq.\ (\ref{action}) opposite to that of
ref.\ \cite{CGHS}. This choice of the sign is convenient for a
semiclassical analysis in the limit $c \rightarrow - \infty$.
The partition function on the genus $h$ Riemann surface is given
by a path integral
\be
Z_\chi(\mu) = \int {\D_g \, g_{\alpha\beta} \D_g \phi \,
\D_g f \over V_{\rm gauge}} \e^{- S}
\label{partition}
\ee
where $\chi = 2 - 2h$ is the Euler number of the surface and
$V_{\rm gauge}$ is the volume of the group of diffeomorphisms.
The functional measures are defined in a diffeomorphism invariant
way using the metric $g_{\alpha\beta}$. In particular, the
measures of the metric and the dilaton are defined by the
norms \cite{POLYAKOV}
\ba
|| \delta g_{\alpha\beta} ||_g^2 \A = \A \int d^2 \xi \sqrt{g} \,
(g^{\alpha\gamma} g^{\beta\delta}
+ u g^{\alpha\beta} g^{\gamma\delta})
\delta g_{\alpha\beta} \delta g_{\gamma\delta} \quad
\left( u > -{1 \over 2} \right), \nonu
|| \delta \phi ||_g^2 \A = \A \int d^2 \xi \sqrt{g} \,
(\delta \phi)^2.
\label{measure}
\ea
In ref.\ \cite{STR} a different choice of the measures was used.
We will discuss such modifications of the measures at the end of
the paper. For a semiclassical analysis it is more convenient to
consider the partition function with a fixed `area' $A$
\be
\tilde Z_\chi(A) = \int {\D_g g_{\alpha\beta} \, \D_g \phi \,
\D_g f \over V_{\rm gauge}} \,
\delta \left( \int d^2 \xi \sqrt{g} e^{-2\phi} - A \right)
\e^{- S}.
\label{apartition}
\ee
The partition function (\ref{partition}) is obtained from
eq.\ (\ref{apartition}) by integrating over $A$.
\par
To fix the diffeomorphism invariance we choose the conformal gauge
\cite{POLYAKOV}
\be
g_{\alpha\beta}(\xi)
= \e^{2\rho(\xi)} \hat g_{\alpha\beta}(\xi; \tau),
\label{gauge}
\ee
where $\hat g_{\alpha\beta}$ is a reference metric which
depends on the moduli $\tau$ of the Riemann surface.
After the gauge fixing and an integration of the matter field
we obtain
\be
\tilde Z_\chi(A)
= \int {[d\tau] \over V_{\rm CKV}} \int \D_g \rho \D_g \phi \,
\delta \left( \int d^2 \xi \sqrt{\hat g} e^{2(\rho-\phi)}
- A \right) \e^{- S_{\rm eff}},
\label{gaugefixed}
\ee
where $V_{\rm CKV}$ is the volume of the group generated by the
conformal Killing vectors, which exist for genera $h = 0, 1$.
The effective action is given by
\be
S_{\rm eff} = {1 \over 2\pi} \int d^2\xi \sqrt{\hat g} \, \left[
\e^{-2\phi} \left( R_{\hat g} - 4 \hat g^{\alpha\beta}
\partial_\alpha \phi \partial_\alpha \rho
+ 4 \hat g^{\alpha\beta} \partial_\alpha \phi \partial_\beta \phi
\right) \right] + S_{\rm L} + {\mu \over \pi} A,
\label{effaction}
\ee
where we have used the fixed area condition. The Liouville action
\be
S_{\rm L}[\hat g_{\alpha\beta}, \rho]
= {\gamma \over 2\pi} \int d^2 \xi \sqrt{\hat g}
\left( \hat g^{\alpha\beta} \partial_\alpha \rho
\partial_\beta \rho + R_{\hat g} \, \rho
+ 2 \mu' \e^{2\rho} \right), \quad
\gamma = {26-c \over 12}
\label{liouville}
\ee
is a result of the Weyl anomaly of the matter and the
Faddeev-Popov ghost fields \cite{POLYAKOV}. The parameter $\mu'$
is regularization dependent and we choose it to be zero for
simplicity.
\par
The functional measures of $\rho$ and $\phi$ in
eq.\ (\ref{gaugefixed}) are defined by the norms induced from
eq.\ (\ref{measure})
\ba
|| \delta \rho ||_g^2 \A = \A \int d^2 \xi \sqrt{\hat g}
\e^{2\rho} (\delta \rho)^2, \nonu
|| \delta \phi ||_g^2 \A = \A \int d^2 \xi \sqrt{\hat g}
\e^{2\rho} (\delta \phi)^2.
\label{rhomeasure}
\ea
Due to the factor $\e^{2\rho}$ in these norms it is not obvious how
to evaluate the functional integral.
In refs.\ \cite{STR,ABC} a relation of these measures to those without
the factor $\e^{2\rho}$ was given as an ansatz following
the procedure in ref.\ \cite{DDK}.
We will check this ansatz by evaluating the functional integral
(\ref{gaugefixed}) in a semiclassical approximation,
which becomes exact for $c \rightarrow - \infty$.
\par
%
%
As a first step of the semiclassical quantization let us find
a saddle point of the functional integral (\ref{gaugefixed}).
We have to find a minimum of the exponent $-S_{\rm eff}$
under the condition of fixed area.
Introducing the Lagrange multiplier $\lambda \in \R$
it can be found by the variational principle of $S_{\rm eff}
+ {\lambda \over \pi} (\int d^2 \xi \sqrt{\hat g}
e^{2(\rho-\phi)} - A)$. The Euler-Lagrange equations are
\ba
2 \Delta_{\hat g} \rho - R_{\hat g} - 2 \lambda \e^{2\rho}
- 4 ( \Delta_{\hat g} \phi - \hat g^{\alpha\beta}
\partial_\alpha \phi \partial_\beta \phi ) \A = \A 0, \nonu
2 \Delta_{\hat g} \rho - R_{\hat g} - {4 \lambda \over \gamma}
\e^{2(\rho-\phi)} - {4 \over \gamma} ( \Delta_{\hat g} \phi
- \hat g^{\alpha\beta} \partial_\alpha \phi \partial_\beta \phi )
e^{-2 \phi} \A = \A 0, \nonu
\int d^2 \xi \sqrt{\hat g} e^{2(\rho-\phi)} - A \A = \A 0.
\label{eleq}
\ea
If we choose the dilaton field to be a constant
\be
\phi_{\rm cl} = -{1 \over 2} \ln {\gamma \over 2},
\label{classphi}
\ee
then the first two equations of eq.\ (\ref{eleq}) are reduced to
a single equation
\be
R_{\bar g} = - 2 \lambda, \quad
\bar g_{\alpha\beta} \equiv \e^{2 \rho_{\rm cl}} \,
\hat g_{\alpha\beta},
\label{classmetric}
\ee
which expresses the fact that the metric $\bar g_{\alpha\beta}$
has a constant curvature.
The multiplier $\lambda$ can be fixed by integrating
eq.\ (\ref{classmetric}) over the surface and using the third
equation of eq.\ (\ref{eleq})
\be
\lambda = - {\pi \gamma \chi \over A}.
\label{lambdavalue}
\ee
It is a mathematical theorem that there exists a unique metric
with a constant curvature on any Riemann surface.
Thus eq.\ (\ref{classmetric}) with eq.\ (\ref{lambdavalue})
determines $\rho_{\rm cl}$.
It is easy to see that the $A$-dependence of $\rho_{\rm cl}$ is
\be
\rho_{\rm cl} = \rho_{\rm cl}^{A=1} + {1 \over 2} \ln A.
\label{adepofrho}
\ee
\par
If we considered the partition function (\ref{partition}) instead
of the area-fixed one (\ref{apartition}), the saddle point
condition would be the first two equations of eq.\ (\ref{eleq})
with $\lambda$ replaced by the cosmological constant $\mu$. Then,
as one can easily see by integrating the equation corresponding
to eq.\ (\ref{classmetric}), a solution of the form
(\ref{classphi}), (\ref{classmetric}) is possible only for either
$\chi > 0$ or $\chi < 0$ depending on the sign of $\mu$.
This is the reason why we have considered the area-fixed
partition function (\ref{apartition}).
\par
%
%
Next we define quantum fluctuations around the above configuration
\be
\tilde \phi = \phi - \phi_{\rm cl}, \quad
\tilde \rho = \rho - \rho_{\rm cl}
\label{fluc}
\ee
and expand the action $S_{\rm eff}$ in these fluctuations.
The action up to quadratic terms is
\ba
S_{\rm eff} \A = \A {\mu \over \pi} A + \gamma\chi
+ S_{\rm L}[\hat g_{\alpha\beta}, \rho^{A=1}_{\rm cl}]
+ \gamma \chi \ln A
+ {\gamma \over 2\pi} \int d^2\xi \sqrt{\bar g} \, \Bigl[ \,
\bar g^{\alpha\beta} \partial_\alpha \tilde \phi
\partial_\beta \tilde \phi \nonu
\A\A + \, {2\pi\gamma\chi \over A} \, \tilde \phi^2
+ \bar g^{\alpha\beta} \partial_\alpha
(\tilde \rho - \tilde \phi) \partial_\beta
(\tilde \rho - \tilde \phi) - {2\pi\gamma\chi \over A} \,
(\tilde \rho - \tilde \phi)^2 \, \Bigr].
\label{expansion}
\ea
To obtain eq.\ (\ref{expansion}) we have used the fixed area
condition
\ba
0 \A = \A \int d^2 \xi \sqrt{\hat g} \e^{2(\rho-\phi)} - A \nonu
\A = \A \gamma \int d^2 \xi \sqrt{\bar g} \left[ \,
(\tilde\rho - \tilde\phi) + (\tilde\rho - \tilde\phi)^2
+ \cdots \, \right]
\label{linear}
\ea
to eliminate linear terms in the expansion.
\par
We now substitute eq.\ (\ref{expansion}) into
eq.\ (\ref{gaugefixed}) and evaluate the integrals.
We decompose the fields into zero (constant) modes
$\rho_0,\ \phi_0$ and nonzero modes $\rho',\ \phi'$
($\tilde\rho = \rho_0 + \rho',\ \tilde\phi = \phi_0 + \phi'$)
with respect to the Laplacian $\Delta_{\bar g}$.
To the one-loop order the norm of $\rho$ (\ref{rhomeasure})
can be approximated as \cite{ZAM,CKT}
\ba
|| \delta \rho ||_g^2 \A = \A \int d^2 \xi \sqrt{\bar g}
\e^{2\tilde\rho} (\delta \tilde\rho)^2, \nonu
\A \cong \A \int d^2 \xi \sqrt{\bar g} \,
(\delta \tilde\rho)^2, \nonu
\A = \A (\delta \rho_0)^2 \int d^2 \xi \sqrt{\bar g}
+ \int d^2 \xi \sqrt{\bar g} \, (\delta \rho')^2.
\label{oneloopmeasure}
\ea
Therefore the functional measure becomes
\be
\D_g \rho \cong d \rho_0 \D_{\bar g} \rho' \left( \int d^2 \xi
\sqrt{\bar g} \right)^{1 \over 2}
= d \rho_0 \D_{\bar g} \rho' \left( {2A \over \gamma}
\right)^{1 \over 2}.
\label{decompmeasure}
\ee
A similar formula holds for the measure of $\phi$.
The $\rho_0$ integral can be evaluated as
\be
\int d \rho_0 \, \delta \left( \int d^2 \xi \sqrt{\hat g}
\e^{2(\rho-\phi)} - A \right) \e^{2\gamma\chi (\rho_0 - \phi_0)^2}
\cong {1 \over 2A},
\label{zeromodeint}
\ee
while the $\phi_0$ integral gives an $A$-independent constant.
The integrals of the nonzero modes $\rho',\ \phi'$ give functional
determinants of differential operators
\be
{\rm det}' \left( \Delta_{\bar g}
- {2\pi\gamma\chi \over A} \right)^{-{1 \over 2}}
{\rm det}' \left( \Delta_{\bar g}
+ {2\pi\gamma\chi \over A} \right)^{-{1 \over 2}},
\label{determinant}
\ee
where the primes denote an exclusion of the zero mode of
$\Delta_{\bar g}$. They are the same type of operators as those
appearing in the semiclassical analysis of the ordinary Liouville
theory \cite{ZAM,CKT}. The $A$-dependence is given
by \cite{ZAM,CKT}
\be
A^{1 \over 2} {\rm det}' \left( \Delta_{\bar g}
+ {4\pi\eta\chi \over A} \right)^{-{1 \over 2}}
= {\rm const.} \times A^{{1 \over 12}\chi
+ {1 \over 2}\eta\chi} \e^{-KA},
\label{det}
\ee
where $K$ is a regularization dependent constant, which we choose
to be zero.
\par
Collecting the above results we finally obtain
the $A$-dependence of the area-fixed partition function
\be
\tilde Z_\chi(A) = {\rm const.} \times
A^{\Gamma(\chi) - 3} \e^{- {\mu \over \pi} A},
\label{adep}
\ee
where the string susceptibility is given by
\be
\Gamma(\chi) = {c - 24 \over 12} \, \chi + 2 + O((-c)^{-1}).
\label{susceptibility}
\ee
The $\mu$-dependence of the partition function (\ref{partition})
is obtained from eq.\ (\ref{adep}) by integrating over $A$
\be
Z_\chi(\mu) = {\rm const.} \times \mu^{-\Gamma(\chi) + 2}.
\label{mudep}
\ee
\par
%
%
Let us compare this result with that of ref.\ \cite{ABC}.
It was argued in ref.\ \cite{ABC} that the functional
integral (\ref{partition}) can be reduced to
\be
Z_\chi(\mu) = \int {[d\tau] \over V_{\rm CKV}}
\int \D_{\hat g} \, \chi \D_{\hat g} \Omega \, \e^{- S_{\rm CFT}},
\label{abc}
\ee
where the action is given by
\be
S_{\rm CFT} = {1 \over 8\pi} \int d^2\xi \sqrt{\hat g} \left(
\hat g^{\alpha\beta} \partial_\alpha \chi \partial_\beta \chi
+ 2 \sqrt{\kappa} R_{\hat g} \, \chi - \hat g^{\alpha\beta}
\partial_\alpha \Omega \partial_\beta \Omega + 2 \mu \kappa
\e^{{1 \over \sqrt{\kappa}}(\chi - \Omega)-1} \right).
\label{abcaction}
\ee
The parameter $\kappa$ is determined by the conformal invariance
as $\kappa = {24-c \over 12}$. To make the integral of $\Omega$
convergent one has to integrate it along the imaginary direction
$\Omega = i \bar \Omega$ ($\bar \Omega \in \R$).
The $\mu$-dependence of the partition function can be found by
shifting the field as $\chi \rightarrow \chi
- \sqrt{\kappa}\ln\mu$ in eq.\ (\ref{abc}).
Thus we find the string susceptibility in this approach
\be
\Gamma_{\rm CFT}(\chi) = {c - 24 \over 12} \, \chi + 2,
\label{abcmudep}
\ee
which is consistent with our semiclassical result
(\ref{susceptibility}).
\par
%
%
Finally let us discuss other choices of the functional measures
in the path integral. In ref.\ \cite{STR} the measures of the
metric and the dilaton were defined using the modified metric
$\e^{-2\phi} g_{\alpha\beta}$, while that of the matter field was
defined by $g_{\alpha\beta}$.
More generally, one may consider also a modification of the matter
measure. Here, we consider the following general situation.
Suppose that among the matter-ghost system contributing to the
Weyl anomaly proportional to $\gamma$, a fraction
${\gamma_i \over \gamma}$ ($i = 1,\ 2,\ 3,\ \cdots,\ \sum_i
\gamma_i = \gamma$) has a measure defined by a modified metric
$\e^{2 \alpha_i \phi} g_{\alpha\beta} = \e^{2(\rho+\alpha_i \phi)}
\hat g_{\alpha\beta}$.
The Liouville action (\ref{liouville}) is then replaced by
\ba
S'_{\rm L} \A = \A \sum_i {\gamma_i \over \gamma} \,
S_{\rm L}[\hat g_{\alpha\beta}, \rho + \alpha_i \phi] \nonu
\A = \A {1 \over 2\pi} \int d^2 \xi \sqrt{\hat g}
\Bigl( \gamma \, \hat g^{\alpha\beta} \partial_\alpha \rho
\partial_\beta \rho + \gamma R_{\hat g} \, \rho
+ a \, \hat g^{\alpha\beta} \partial_\alpha \phi
\partial_\beta \phi \nonu
\A\A + \, 2 b \, \hat g^{\alpha\beta} \partial_\alpha \phi
\partial_\beta \rho + b \, R_{\hat g} \, \phi \Bigr),
\label{mliouville}
\ea
where $a = \sum_i \gamma_i \alpha_i^2,\
b = \sum_i \gamma_i \alpha_i$.
The case we have considered so far corresponds to $a = b = 0$,
while the choice in ref.\ \cite{STR} corresponds to
$a = -b = {13 \over 6}$. The measures of $\rho$ and $\phi$
(\ref{rhomeasure}) may also be modified. However, it does not
change the result of the string susceptibility in the
one-loop approximation.
\par
Using eq.\ (\ref{mliouville}) instead of eq.\ (\ref{liouville})
we can repeat the above semiclassical analysis.
The saddle point is found to be
\be
\phi_{\rm cl} = -{1 \over 2} \ln {\gamma + b \over 2}, \quad
R_{\bar g} = {2 \pi (\gamma+b) \chi \over A},
\label{mclassfield}
\ee
where $\bar g_{\alpha\beta}$ is defined in
eq.\ (\ref{classmetric}).
Integrating the fluctuations around the configuration
(\ref{mclassfield}) we find that the area-fixed partition function
is given by the same expression as before except that the
functional determinants (\ref{determinant}) are replaced by
\be
{\rm det}' \left( \Delta_{\bar g} - {2\pi(\gamma+b)^2\chi \over
(\gamma + a + 2b) A} \right)^{-{1 \over 2}}
{\rm det}' \left( \Delta_{\bar g} + {2\pi(\gamma+b)\chi \over A}
\right)^{-{1 \over 2}}.
\label{mdeterminant}
\ee
Thus we find the string susceptibility
\be
\Gamma(\chi) = \left[ {c - 24 \over 12}
+ {(a+b) (26-c+12b) \over 4 (26-c+12a+24b)} \right] \chi
+ 2 + O((-c)^{-1}).
\label{msusceptibility}
\ee
The choice in ref.\ \cite{STR} $a = -b = {13 \over 6}$ gives the
same susceptibility as eq.\ (\ref{susceptibility}),
which is consistent with the result of refs.\ \cite{STR,ABC}.
It is interesting to study the model with other choices of $a$
and $b$ using the method of refs.\ \cite{STR,ABC} and compare its
result with eq.\ (\ref{msusceptibility}).
\par
\vspace{5mm}
The author would like to thank S. Yamaguchi for a collaboration in
an early stage of this work. He would also like to thank N. Sakai,
Y. Matsumura and T. Uchino for useful discussions on the dilaton
gravity.
%
%

%

\newpage
\begin{thebibliography}{100}
%
\bibitem{CGHS} C.G. Callan, S.B. Giddings, J.A. Harvey and
        A. Strominger, {\it Phys.\ Rev.\ }{\bf D45} (1991) R1005.
\bibitem{VERLINDE} H. Verlinde,
        in {\it String Theory and Quantum Gravity '91}, eds.\
        J. Harvey et al., (World Scientific, Singapore, 1992).
\bibitem{STR} A. Strominger, Santa Barbara preprint UCSBTH--92--18
        (May, 1992).
\bibitem{ABC} S.P. de Alwis, Colorado preprint COLO--HEP--280
        (May, 1992); A. Bilal and C. Callan, Princeton preprint
        PUPT--1320 (May, 1992).
\bibitem{RT} J.G. Russo and A.A. Tseytlin, Stanford and DAMTP
        preprint SU--ITP--92--2, DAMTP--1--1992 (January, 1992);
        K. Hamada, Tokyo preprint UT--Komaba 92--7 (June, 1992);
        A. Mikovi\'c, Queen Mary preprint QMW/PH/92/12
        (June, 1992).
\bibitem{DDK} F. David, {\it Mod.\ Phys.\ Lett.\ }{\bf A3}
        (1988) 1651; J. Distler and H. Kawai,
        {\it Nucl.\ Phys.\ }{\bf B321} (1989) 509.
\bibitem{KM} V.A. Kazakov and A.A. Migdal,
        {\it Nucl.\ Phys.\ }{\bf B311} (1988/89) 171.
\bibitem{ZAM} A.B. Zamolodchikov,
        {\it Phys.\ Lett.\ }{\bf B117} (1982) 87.
\bibitem{CKT} S. Chaudhuri, H. Kawai and S.-H.H. Tye,
        {\it Phys.\ Rev.\ }{\bf D36} (1987) 1148.
\bibitem{KPZ} V.G. Knizhnik, A.M. Polyakov and A.B. Zamolodchikov,
        {\it Mod.\ Phys.\ Lett.\ }{\bf A3} (1988) 819.
\bibitem{MMDK} N.E. Mavromatos and J.L. Miramontes,
        {\it Mod.\ Phys.\ Lett.\ }{\bf A4} (1989) 1847;
        E. D'Hoker and P.S. Kurzepa,
        {\it Mod.\ Phys.\ Lett.\ }{\bf A5} (1990) 1411;
        E. D'Hoker, {\it Mod.\ Phys.\ Lett.\ }{\bf A6} (1991) 745.
\bibitem{POLYAKOV} A.M. Polyakov, {\it Phys.\ Lett.\ }{\bf B103}
        (1981) 207; {\it Gauge Fields and Strings},
        (Harwood Academic Publishers, Chur, 1987).
%
\end{thebibliography}
\end{document}